\documentstyle[12pt,twoside]{article}
\def\be{\begin{equation}}
\def\ee{\end{equation}}
\def\bn{\begin{eqnarray}}
\def\en{\end{eqnarray}}

\def\n{\noindent}
\def\f{\phi}
\def\h{{\phi}_{,r}}

\def\p0{{\phi}_{0}}
\def\g{g_{,r}}

\def\crr{g_{,r,r}}
\def\tr{T_{,r}}

\def\rr{R_{,r}}
\def\pt{R_{,t}}
\def\qt{T_{,t}}
\def\dr{D_{,r}}
\def\se{D_{,t,t}}
\def\arr{D_{,r,r}}
\def\bh{black hole}
\def\bhs{black holes}
\def\sdk{Schwarzschild-like}
\def\srk{Schr\"{o}dinger-like}
\def\ger{Schr\"{o}dinger}
\def\2d{wo-dimensional}
\def\m{metric}
\def\e{{\rm e}}
\def\d{{\rm d}}
\def\to{\rightarrow}
\def\gen{eigenvalue}
\def\genr{general relativity}
\def\gere{general-relativity}
\def\alf{{\textstyle\frac{1}{4}}}

\begin{document}
\renewcommand{\thefootnote}{\fnsymbol{footnote}}
\begin{center}

{\Large\bf Instability of t\2d\ heterotic \\[5pt]
           stringy \bhs}
\medskip

{\bf Mustapha Azreg-A\"{\i}nou\footnote[1]{E-mail address:
azreg.as@mozart.emu.edu.tr, azreg@hermes.unice.fr}}
\medskip

Eastern Mediterranean University, Department of Physics,\\
Gazimagusa, North Cyprus,\\ (Via Mersin 10, Turkey)

\end{center}

\begin{abstract}
We solve the \gen\ problem of \genr\ for the case of charged \bhs\ in t\2d\
heterotic string theory, derived by McGuigan {\it et al\/}. For the case of
$m^{2}>q^{2}$, we find a physically acceptable time-dependent growing mode;
thus the \bh\ is unstable. The extremal case $m^{2}=q^{2}$ is stable.
\end{abstract}

\section{Introduction}
T\2d\ low-energy string theory admits several \bh\ solutions. Some of them
are neutral, like the \bh\ solution to the bosonic closed string theory on
the sphere \cite{witten,mandal,steif} and \cite{callan}. The others are
charged \cite{yost}, open string Born-Infeld \bhs\ or heterotic string \bhs.
These two classes of charged \bhs\ correspond to two different alternative
ways to couple a gauge field to string theory (with a non-minimal coupling of the
scalar dilaton field to gravity). They have very interesting and complex
spacetime geometries with more than two horizons depending on the number of
string loop corrections included \cite{yost}.

Their mass, charge, temperature and entropy have been evaluated in different
ways \cite{yost,gibbons,nappi}. In \cite{gibbons}, and then elegantly in
\cite{nappi}, the authors used thermodynamic arguments, but different methods,
to compute the entropy and other thermodynamic quantities. While, recently,
Teo \cite{teo} has extended the work of Hyun \cite{hyun}, and by Sfetsos
\& Skenderis \cite{sfetsos}, and derived the entropy for the t\2d\ \bh\ by
explicitly establishing the U-duality between the t\2d\ \bh\ and the
five-dimensional one. Very recently, a similar work \cite{9806036}, calling
on, this time, a sequence of S and T-S-T duality transformations in four
dimensions, has been performed leading to the same expressions of the entropy
for t\2d\ \bhs.

It is now obvious that the thermodynamics of higher-dimensional \bhs\ may be
described by that of lower-dimensional \bhs\ and vice versa. And one
question arises: Is this observation extendible to other physical properties?
In other words: May the much simpler physics of lower-dimensional \bhs\
describe faithfully that of higher-dimensional \bhs? Suggestions that
higher-dimensional \bhs\ might be governed by the same conformal field
theories as lower-dimensional ones have been put forward
\cite{teo,larsen,satoh}.

Studying the propagation of scalars, Satoh \cite{satoh} has shown some common
features between higher- and lower-dimensional \bhs. Almost similar studies
carried out in \cite{holzhey} for four-dimensional Einstein-Maxwell-Dilaton 
\bhs\ (without potential for the dilatonic field) and in \cite{cold} for
four-dimensional cold scalar-tensor \bhs, respectively, showed that for some
arrangement of the parameters (infinite temperature and vanishing entropy),
the \bhs\ even become repulsive. We will prove almost similar properties for
t\2d\ \bhs\ \cite{non}. In \cite{holzhey}, the authors went further in their
conclusions and argued that certain classes of four-dimensional Einstein-
Maxwell-Dilaton \bhs\ behave as follows: Some, like extended objects (liquid
drop), exhibit time delay for the re-emission, and others, like elementary
particles, exhibit no time delay for the re-emission of low-energy quanta.
But, before an interpretation in terms of liquid drop or elementary particles
can be achieved, and before any further development, the condition of
stability must be fulfilled.

In this paper we address the question of stability of a class of t\2d\
stringy \bhs. We mention that the small fluctuations of the Witten
solution \cite{witten} and other solutions have been studied by some authors
who gave no conclusion regarding the stability of the solutions investigated
\cite{with}. While in \cite{diamandis} the authors, investigating the
linearized time-dependent perturbations of t\2d\ stingy \bhs\ in the presence
of real tachyon field, concluded that there is no time-dependent solutions
with horizons shrinking to a point; rather they expand with increasing time,
and are thus unstable solutions. For the first time, to our knowledge, the
stability of t\2d\ heterotic stringy \bhs\ has been studied, so far,
explicitly by Hsu {\it et al} \cite{hsu}. The authors' analysis, which we
believe to be flawed, does not deal with the problem of stability as it is
defined in \genr, especially in the presence of dilatonic field.

Motivated by all these developments and analyses mentioned above, we intend
to re-examine the case of heterotic stringy \bhs\ by completing their
stability analysis, on the one hand, and clarifying the \gen\ problem of
\genr\ on the other hand. The paper is organized as follows. In section 2 we
review the \bh\ heterotic string solution. In section 3 we write the
linearized equations governing the evolution of small perturbations and the
resulting \srk\ equation for a specific function related to the small
perturbations, together with some useful relationships between them. In
section 4 we define the \gere\ \gen\ problem corresponding to the stability
analysis, which is, in general, different from the \ger\ \gen\ problem
obtained in section 3. We solve the former and draw a parallel with the
stability investigation given by Hsu {\it et al} \cite{hsu}.

\section{Black holes in t\2d\ heterotic string theory}
The t\2d\ \m\ of heterotic stringy \bh\ can be parametrized by \cite{yost}

\be
\label{m}
\d s^{\,2} = g(r)\,\d t^2 - \frac{1}{g(r)}\,\d r^2
\ee

\n with

\be
g(r) = 1-2\,m\,\e ^{-Q\,r} + q^2\,\e ^{-2\,Q\,r}
\ee

\n the dilaton field background is that one of a neutral stringy \bh,

\be
\f (r) = \p0 - \frac{Q}{2}\,r
\ee

\n while the gauge field $F_{tr} \equiv f$ reads

\be
\label{g}
f(r) = \sqrt{2}\,Q\,q\,\e ^{-Q\,r}\,.
\ee

\n The {\m}, dilatonic and gauge fields given by equations (\ref{m} $\to$
\ref{g}) are solutions to the equations of motion

\bn
R_{\mu\nu} - 2\,\f _{;\mu ;\nu} - \textstyle\frac{1}{2}\,F_{\mu\rho}\,
F_{\nu}^{\,\rho} = 0 \nonumber \\
\label{f}
\left(\e ^{-2\,\f}\,F^{\mu\nu}\right)_{;\nu} = 0 \\
R + 4\,\f _{;\rho}\,\f ^{;\rho} - 4\,\f _{;\rho}^{;\rho} - c - \alf \,
F^{2} = 0 \nonumber 
\en

\n (in our conventions $R_{\mu\nu\rho\sigma}$ is defined so that
$R_{trtr}=-\crr /2$ or, equivalently, $R_{rr}=\crr /2g$), which are derived
from the effective action \cite{yost}

\be
S = \int \d ^{2}\,x\,\sqrt{-G}\,\e ^{-2\,\f}\,\left[R - 4\,(\nabla \f)^{2} - c
- \alf \,F^{2}\right]\,.
\ee

\n The parameters $m$, $q$ are related to the mass, $\cal M$, and the electric
charge, $\cal Q$, respectively, of the \bh\ by

$$
{\cal M} = 2\,Q\,m\,\e ^{-2\,\p0} 
$$
$$
{\cal Q} = \sqrt{2}\,Q\,q\,\e ^{-2\,\p0}
$$

\n while $Q$, $\p0$ are just integration constants, and $c$ is the central
charge \cite{callan}. The sign of $Q>0$ is fixed by the asymptotic flatness
condition at $r=+\infty\,$\footnote{Note that the case $c=0$ is trivial.},
while its value is related to $c$ since the third equation in (\ref{f})
requires

\be
c = - Q^{2}\,.
\ee

\n The entropy and the temperature are given in terms of $\cal M,\,\cal Q$,
respectively by \cite{nappi}

$$
{\cal S} = \frac{2\,\pi}{Q}\,\left({\cal M} + \sqrt{{\cal M}^2 - 2\,{\cal Q}^2}
           \right)
$$
$$
{\cal T} = \frac{Q}{2\,\pi}\,\frac{\sqrt{{\cal M}^2 - 2\,{\cal Q}^2}}
           {{\cal M} + \sqrt{{\cal M}^2 - 2\,{\cal Q}^2}}\,.
$$

\n According to \cite{holzhey}, the condition for the thermal description to
break down is

$$
\left(\frac{\partial{\cal T}}{\partial{\cal M}}\right)_{{\cal Q}} \equiv
\frac{Q}{4\,\pi\,{\cal Q}^2}\,\left[\frac{2\,({\cal M}^2 - {\cal Q}^2)}
{\sqrt{{\cal M}^2 - 2\,{\cal Q}^2}} - 2\,{\cal M}\right] \gg 1
$$

\n is largely fulfilled for the extremal \bh\, $m ^2 = q ^2$
(${\cal M}^2 = 2{\cal Q}^2$); a small change in the mass is accompanied by a
huge change in the temperature of the hole, the thermal description becomes
ambiguous.

\n The curvature singularity is at $r=-\infty$ since $R=-\crr\,$. Putting
${\cal R}(r)=\e ^{-Q\,r}$, the roots of the equation $g({\cal R})=0$ read

\be
{\cal R}_{\pm} = \frac{m \pm \sqrt{m^{2} - q^{2}}}{q^{2}}\,.
\ee

\n There are event horizons (apparent singularities) at $r_{\pm}$ related to
${\cal R}_{\pm}$ by

\be
{\cal R}_{+} = \e ^{-Q\,r_{-}}\;\;\;\;\;\;\;{\cal R}_{-} = \e ^{-Q\,r_{+}}\,.
\ee

\n From now on we shall consider the case $m^{2}\geq q^{2}$, $m>0$ and $q$
real, corresponding to two real values $r_{+}\geq r_{-}>0$ with two event
horizons.

\section{The linearized field equations for small perturbations}
In the standard terminology employed by Chandrasekhar \cite{bh}, the
perturbation functions of the background fields are grouped into two sets,
polar and axial perturbations. In our case, the polar set is that one which
preserves the ``initial" {\em diagonal form} of the background metric
(\ref{m}). Now, for any t\2d\ initial \m\ configuration, the most sufficiently
general form of the associated perturbed \m\ can always be brought to a
diagonal form. The proof is given in the chapter 2 of \cite{bh} in the case of
a \m\ with a positive- or negative-definite signature ($+,+$ or $-,-$) and is
manifestly generalized to our case with a signature ($+,-$). Consequently,
t\2d\ \m\ perturbations are entirely polar. In our case the perturbed metric
can be written as

\be
\label{gm}
\d s^{\,2} = [g(r) + T(r,t)]\,\d t^{\,2} - \left(\frac{1}{g(r)} +
Y(r,t)\right)\,\d r^{\,2}
\ee

\n and the perturbations of the dilatonic and Maxwell fields are introduced by

\be
\label{gd}
\begin{array}{lll}
\f (r,t) & = & \f (r) + D(r,t)\\
f(r,t) & = & f(r) + M(r,t) \end{array}
\ee

\n where $T$, $Y$, $D$ and $M$ are considered to be ``relatively" small
quantities compared to the background fields. So, this yields an inventory
of four variables.

Notice that it is still possible to fixe the gauge and supply the three
linearized field equations (see Eqs. (\ref{tr} $\to$ \ref{rr}) below) with an
extra equation, which yields an inventory of four equations. As shown in the
appendix, we can always bring the diagonal sufficiently generalized t\2d\ \m\
to a \sdk\ form where $g_{rr}=-1/g_{tt}$. In our case, we have $[g(r)+T(r,t)]
[(1/g(r))+Y(r,t)]=1$. Keeping only linear terms, we get

\be
\label{sdk}
g\,Y + \frac{T}{g} = 0\,.
\ee\

The basic three linearized field equations read\footnote{The Eqs. (\ref{tr}
$\to$ \ref{rr}) have been given in \cite{hsu}, where ``$\partial /\partial t$"
has been replaced by ``$i\sigma$", but without specifying the combinations in
the l.h.s leading to these equations.} (substitute
(\ref{gm}), (\ref{gd}) into (\ref{f}))

\be
\label{tr}
\bar{\ell}_{tr}^{(G)} = g\,\h\,Y - 2\,\dr + \frac{\g}{g}\,D = 0
\;\;\;\;\;,\;\;\;\;\bar{\ell}^{(A)\,r} = - 2\,f\,D + M = 0
\ee
\be
\label{rr}
\ell_{rr} + \frac{\ell^{(\f)}}{2\,g} + \frac{f}{2\,g}\,\ell^{(A)\,r} +
\frac{\g}{g}\,\bar{\ell}_{tr}^{(G)} - 2\,\h\,\bar{\ell}_{tr}^{(G)} -
\bar{\ell}_{tr,r}^{(G)} = -\,\frac{2}{g^{2}}\,\se - 2\,\arr = 0
\ee

\n where equation (\ref{sdk}) has been used\footnote{In Eqs. (\ref{tr} $\to$
\ref{rr}), $\ell_{\mu\nu}^{(G)}$, $\ell^{(A)\,\mu}$, $\ell^{(\f)}$ are
the linear variations of the l.h.s of Eqs. (\ref{f}), respectively, and
$\bar{\ell}_{\mu\nu}^{(G)}$, $\bar{\ell}^{(A)\,\mu}$, $\bar{\ell}^{(\f)}$
are their primitives with respect to the time coordinate.}.

In terms of the tortoise $r^{*}$-coordinate, the equation (\ref{rr})  
can be brought to \srk\ equation via the transformations $\d r^{*}\equiv
(1/g(r))\,\d r$, $\psi(r^{*},t)\equiv D/\sqrt{g}\,$. This yields \cite{hsu}

\be
\label{ann}
-\psi_{,r^{*},r^{*}} + \left(\frac{3}{4\,g^{2}}\,(g_{,r^{*}})^{2} -
\frac{1}{2\,g}\,g_{,r^{*},r^{*}}\right)\,\psi = \psi_{,t,t}
\ee

\n where the function between round brackets is the potential $V(r^{*})$. It
can be put on the following form which we shall use in section 4

\be
\label{v}
V = \frac{1}{2}\,\left(\frac{1}{2}\,(\g)^{2} - g\,\crr\right)\,.
\ee

\section{Stability analysis}
Starting with small perturbations of the background fields at $t=0$, we
examine if they may increase indefinitely in time. For stationary
perturbations of the form $D(r,t)=D(r)$e$^{i\,\omega \,t}$, etc, this means
searching for physically acceptable solutions to the linearized field
equations for $\omega$ imaginary, ie $k^{2} \equiv -\omega ^{2}$. If such
solutions exit, then an initially small perturbation grows exponentially
in time, and the background solution is unstable. Conversely, the solution
is stable if all the eigenvalues $\omega$ are real.

Imposing that the perturbed fields should be regular is a natural
definition of physically acceptable perturbations of regular background
fields. This is not the case of the static component $g_{rr}$ of the \m\
which diverges at $r=r_{\pm}$. In this case, we suppose that the relative
perturbations $T(r)/g(r),\,D(r)/\f (r),\,M(r)/f(r)$ and $Y(r)g(r)=(Y/(1/g))$
are bounded \cite{julio,kk} in the range $]r_{+},\,+\infty[$, which is the
physically interesting region

\be
|gY|=|T/g|<\infty\;\;\;\;\;\;|D/\f|<\infty\;\;\;\;\;\;|M/f|<\infty\;\;\;\;\;\;
(r\in ]r_{+},\,+\infty[)\,.
\ee

\n This minimal requirement --referred to as the weak boundary condition in
\cite{julio}-- defines the \gere\ \gen\ problem , which is, as we shall see,
not necessarily equivalent \cite{az} to the requirement that the
{\em auxiliary} function $\psi$, the solution to equation (\ref{ann}), should
be square integrable in the range $]r_{+},\,+\infty[$, which is the \ger\
\gen\ problem.

For a time dependence of the form $T(r,t)=T(r)$e$^{i\,\omega \,t}$,
$Y(r,t)=Y(r)$e$^{i\,\omega \,t}$, $D(r,t)=D(r)$e$^{i\,\omega \,t}$,
$M(r,t)=M(r)$e$^{i\,\omega \,t}$, with $\omega=-ik$ and $k>0$, the three
equations (\ref{tr}), (\ref{rr}) take the form

\bn
\label{1}
g\,Y & = & \frac{1}{\f _{,\rho}}\,\left(2\,D_{,\rho} - \frac{g_{,\rho}}{g}\,
D\right) \\
\label{2}
\frac{M}{f} & = & 2\,D \\
\label{3}
D_{,\rho,\rho} & + & \frac{k^{2}}{g^{2}}\,D \,= \,0
\en

\n where the new variable $\rho \in ]0,\,+\infty[$ is a translation of
$r:\,\rho\equiv r-r_{+}$.

For \underline{$m^{2}>q^{2}$}, consider the case where

\be
\label{val}
k = k_{0} \equiv \sqrt{m^{2} - q^{2}}\,Q\,{\cal R}_{-}
\ee

\n (it is straighforward to show that $k_{0}=2\pi{\cal T}$). The general
solution to the equation (\ref{3}) can be put into the form

\be
\label{s3}
D(\rho) = D_{0}(\rho)\,\left(a + b\,\int^{\rho}\frac{\d \rho '}
{D_{0}^{2}(\rho ')}
\right)
\ee

\n where $a,\,b$ are real constants. The function $D_{0}(\rho)$ is a
particular solution to the equation (\ref{3}) developed by

\be
\label{d1}
D_{0}(\rho) \equiv \sqrt{\rho}\,\left(1 + \frac{L}{2}\,\rho + \cdots \right)
\ee

\n where $L\equiv Q^{2}(3m{\cal R}_{-}-2)/(2k_{0})$ and the function between
round brackets is an integer series in $\rho$.

The asymptotic behaviour for $\rho \to +\infty$ of the function $D(\rho)$
can be obtained by directly considering the equation (\ref{3}), while its
behaviour at $\rho =0_{-}$ can be deduced from equation (\ref{s3}). For
$\rho \to +\infty$, equation (\ref{3}) reads

\be
D_{,\rho,\rho} + k_{0}^{2}\,D \simeq 0
\ee

\n which is solved by

\be
\label{as}
D(\rho) \simeq c_{1}\,\cos k_{0}\rho + c_{2}\,\sin k_{0}\rho 
\ee

\n where $c_{1},\,c_{2}$ are two arbitrarily real constants. This {\em
harmonic} behaviour for $\rho \to +\infty$ is transmitted via equations
(\ref{2}), (\ref{1}) and (\ref{sdk}) to the relative perturbations $M/f,\,
gY=-T/g$. Hence, the relative perturbations are bounded at $\rho =+\infty$.\\
Substituting (\ref{d1}) into (\ref{s3}) and keeping only up to the first
powers in $\rho$, we get for $\rho \to 0_{-}$

\be
\label{dho}
D(\rho) \simeq \sqrt{\rho}\,\left(a + \frac{(a - 2\,b)}{2}\,L\,\rho + \cdots
\right) + \sqrt{\rho}\,\left(b + \frac{b}{2}\,L\,\rho + \cdots \right)\,
\ln \rho
\ee

\n which is bounded. Upon substituting this function into equation (\ref{1}),
we obtain the following expression for $gY=-T/g$ as $\rho \to 0_{-}$

\be
\label{q}
g\,Y(\rho) \simeq -\frac{2}{Q}\,\frac{b}{\sqrt{\rho}}\,(2 - L\,\rho + \cdots)
\ee

\n which diverges, while the ratio $M/f$ ($=2D$) remains bounded. To deal with
this situation we have to choose $b=0$ in equation (\ref{dho}), which is
possible by suitably fixing the ratio $c_{1}/c_{2}$ in equation (\ref{as}).

Hence, for $k=k_{0}$ and for a suitable value of the ratio $c_{1}/c_{2}$,
not only are the relative perturbations $D/\f$, $M/f$, $gY$, $T/g$ bounded at
both $\rho =0_{-}\,,\;\rho =+\infty$ but the perturbations themselves as
well. These perturbations are also bounded everywhere in the range
$]0,\,+\infty[$ of the variable $\rho$. This can be shown by developing the
perturbations around any point $\rho _0\in ]0,\,+\infty[$. Since the
background fields $g$, $\f$, $f$ are regular functions for all
$\rho \in ]0,\,+\infty[$, the general behaviour --depending on two arbitrary
constants-- of the perturbations given by the Frobenius Method is bounded in
the neighbourhood of $\rho _0$. This mode of perturbation is physically
acceptable. Since it grows in time like $\e ^{k _0\,t}$, the static background
\bh\ solution is unstable. The function $D_{0}$ with the corresponding
functions $T_{0},\,Y_{0},\,M_{0}$ and the \gen\ $k_{0}$ constitute the
solution to the \gere\ \gen\ problem.

The case \underline{$m^{2}=q^{2}$} is quite different. For $\rho \to 0_{-}$,
the equation (\ref{3}) behaves as

\be
\label{m=q}
D_{,\rho,\rho} + \frac{k^2}{Q^{4}\,\rho^4}\,D \simeq 0\,.
\ee

\n The general solution to the exact equation --replacing ``$\simeq$" by
``$=$"-- is given by the r.h.s of the following equation expressing the
behaviour of $D$ when $\rho \to 0_{-}$

\be
\label{sincos}
D(\rho) \simeq a\,\rho\,\cos\left(\frac{k}{Q^{2}\,\rho}\right) + b\,\rho\,
\sin\left(\frac{k}{Q^{2}\,\rho}\right)
\ee

\n ($a,\,b$ are real constants), which is well bounded $\forall\,k>0$, giving
rise to a bounded behaviour of the ratio $M/f$ ($=2D$) but to an
unbounded ratio $gY=-T/g$ $\forall\,a,\,b,\, k>0$, as shown by

\be
g\,Y = -\,\frac{T}{g} \simeq \frac{4}{Q^3}\,\frac{1}{\rho}\,
\left[b\,\cos\left(\frac{k}{Q^{2}\,\rho}\right) -
a\,\sin\left(\frac{k}{Q^{2}\,\rho}\right)\right]\,.
\ee

\n Consequently, there are no physically acceptable growing modes for the
case $m^2 = q^2$; thus it is stable.

Notice that the uncharged case $q=0,\,m>0$, which is a special case of $m^2 >
q^2$, is unstable, for the equations (\ref{s3} $\to$ \ref{q}), as well as
the following discussion, are still valid with, in this situation, $k_0 = Q/2,
\, L=-Q/2$. 

We now consider the \srk\ equation (\ref{ann}). Putting $\psi(r^{*},t)=
\psi(r^{*})\e ^{kt}\,(k>0)$ in the equation, we get

\be
\label{annn}
-\psi_{,r^{*},r^{*}} + (V(r^{*}) - k^{2})\,\psi = 0
\ee

\n where $r^{*}\in ]-\infty,+\infty[$, the horizon $r_{+}$ is at $r^{*}
=-\infty$ and the flat region is at $r^{*}=+\infty$. We can show analytically
that $V(r^{*})>0$ for all $r^{*}\in ]-\infty,+\infty[\,$\footnote{This has
been done numerically in \cite{hsu}.}. In fact, we have $V(r^{*}=-\infty)
=k_{0}^{2}$, $V(r^{*}=+\infty)=0$. From (\ref{v}), we obtain

$$
V_{,r}=-\,\frac{g}{2}\,g_{,r,r,r} = Q ^3\,\left(4\,q^{2}\,\e ^{-Q\,r} -
                                    m\right)\,g(r)\,\e ^{-Q\,r}
$$

\n with $g(r)>0$ for $r_{+}<r<+\infty\,$. Hence, if $16q^{2}/7\geq m^{2}$,
the potential has a maximum at $r_{\rm max}\geq r_{+}$, it is then positive.
If $m^{2}>16q^{2}/7$, the potential is a monotonic function decreasing from
$k_{0}^{2} \to 0$ when $r$ runs from $r_{+} \to +\infty$.

Since $V(r^{*})>0$ for all $r^{*}\in ]-\infty,+\infty[$, \ger\ equation
(\ref{annn}) does not admit bound states (with $k^{2}>0$) vanishing
at both $r^{*}=-\infty ,\,r^{*}=+\infty$. This has led Hsu {\it et al}
\cite{hsu} to the conclusion that the background solution is stable, they
have then studied the \ger\ \gen\ problem for the auxiliary function $\psi$,
and not the \gere\ \gen\ problem we have formulated for the relative
perturbations. We notice that \ger\ equation (\ref{annn}) admits, for all
$0<k^{2}<k_{0}^{2}$, {\em finite} unbound states vanishing exponentially at
$r^{*}=-\infty$ and oscillating at $r^{*}=+\infty$, which are thus non-square
integrable but can be normalized \cite{landau}. It also admits a finite
unbound state for $k^{2}=k_{0}^{2}$ behaving as a constant at $r^{*}=-\infty$
and oscillating at $r^{*}=+\infty$.

\section{Conclusion}
Our analysis, given in section 4, has led to the conclusion that the t\2d\
heterotic sringy \bhs\ are unstable ($m^{2}>q^{2}$), except the extremal case
($m^{2}=q^{2}$), which is stable. For $m^{2}>q^{2}$ (${\cal M}^{2}>
2{\cal Q}^{2}$), we have found a bounded mode, solution to the \gere\ \gen\
problem, growing with time as exp($2\pi{\cal T}t$), where $\cal T$ is the
temperature of the hole. We have seen that the perturbations of the
{\em physical} background fields associated with this mode are {\em all}
bounded; there is then no {\em objection} that this mode is {\em well}
accepted {\em physically\/}. We have also seen that the \gere\ \gen\ problem,
we have solved in section 4, is not equivalent to the \ger\ \gen\ problem.
The latter has been outlined in the last paragraph of section 4 of the present
paper, as well as in \cite{hsu}, where the stability analysis has rested on an
auxiliary function $\psi\,$, whose direct physical meaning is not transparent,
and led to the conclusion that all the \bhs\ ($m^{2}\geq q^{2}$) are stable,
which agrees partially with our conclusion. In a subsequent work, we shall
investigate the stability of all the t\2d\ stringy \bhs.

The case of four-dimensional stringy \bhs\ has been investigated
in \cite{carlini,alwis}. The results of the stability analysis given in
\cite{carlini} are consistent with those of the present case, and different
from those given in \cite{alwis}.

\section*{Appendix}
\appendix
\def\theequation{A.\arabic{equation}}
\setcounter{equation}{0}
To make this article self-contained we show in this appendix how to bring a 
t\2d\ diagonal \m\ to \sdk\ \m. The functions introduced here have nothing
to do with those of the same notation introduced in the text. We start with
the \m\

\bn
\d s^{\,2} & = & g_{tt}(r,t)\,\d t^{\,2} + g_{rr}(r,t)\,\d r^{\,2}
\nonumber \\
\label{a1}
& = & F(r,t)\,\d t^{\,2} - H(r,t)\,\d r^{\,2}
\en

\n and introduce new coordinates $t',\,r'$ by

$$
t' = T(t,r)\;\;\;\;\;\;\;r' = R(t,r)
$$

\n so that the new \m\ preserves its diagonality, $g'_{t'r'}\equiv 0$, and
is \sdk\ \m, $g'_{t't'}g'_{r'r'}=-1$. These two conditions are given
respectively by

\bn
\label{a22}
H\,\pt\,\qt - F\,\rr\,\tr &=& 0\\
\label{a3}
[F\,(\tr)^2 - H\,(\qt)^2]\,[H\,(\pt)^2 - F\,(\rr)^2] &=& {\delta}^4
\en

\n where

\be
\label{a4}
\delta \equiv \qt\,\rr - \tr\,\pt\;(\neq 0)\,.
\ee

\n Developing (\ref{a3}) with the help of (\ref{a22}), we arrive at ${\delta}
^{2}=HF$, or equivalently 

\be
\label{a5}
\qt\,\rr - \tr\,\pt = \pm\,\sqrt{H\,F}\,.
\ee

\n Combining (\ref{a22}), (\ref{a5}) we obtain

\be
\label{a6}
\begin{array}{ccccc}
(H\,\pt)\,.\,\qt &-& (F\,\rr)\,.\,\tr &=& 0 \\
\rr\,.\,\qt &-& \pt\,.\,\tr &=& \pm\,\sqrt{H\,F}\,.
\end{array}
\ee

\n The system (\ref{a6}) is solved with respect to $\qt,\,\tr$ by

\be
\label{a7}
\qt = \pm\,\frac{F\,\sqrt{H\,F}\,\rr}{F\,(\rr)^2 - H\,(\pt)^2}\;\;\;\;\;\;\;
\tr = \pm\,\frac{H\,\sqrt{H\,F}\,\pt}{F\,(\rr)^2 - H\,(\pt)^2}\,.
\ee

\n The integrability condition of the system (\ref{a7}) yields

\be
\label{a8}
\Delta R = \left[\ln |R_{,\rho}\,R^{,\rho}|\right]_{,\mu}\,R^{,\mu}
\ee

\n where

\bn
\Delta R & \equiv & \frac{1}{\sqrt{G}}\,\left(\sqrt{G}\,g^{\mu\nu}\,R_{,\nu}
\right)_{,\mu} \nonumber \\
R^{,\mu} & \equiv & g^{\mu\nu}\,R_{,\nu} \nonumber
\en

\n with $G\equiv |{\rm det}g_{\mu\nu}|=FH$. Similarly, we get by symmetry

\be
\label{a9}
\Delta T = \left[\ln |T_{,\rho}\,T^{,\rho}|\right]_{,\mu}\,T^{,\mu}\,.
\ee

\n It is then possible to bring the \m\ (\ref{a1}) to a diagonal \sdk\ form
by a coordinate transformation $t'=T(t,r),\, r'=R(t,r)$, where $R(t,r),\,
T(t,r)$ are solutions to the differential equation

\be
\label{a10}
\Delta \Psi = \left[\ln |\Psi_{,\rho}\,\Psi^{,\rho}|\right]_{,\mu}\,
\Psi^{,\mu}\,.
\ee

\subsection*{Acknowledgements}
I would like to thank Yost and Carlini for sending copies of
the reprints \cite{yost,carlini}.

\end{document}